\documentclass[10pt,aps,prb,raggedbottom,longbibliography,nobalancelastpage,reprint,citeautoscript,letterpaper,superscriptaddress]{revtex4-2} 
\usepackage[usenames,dvipsnames]{color}
\usepackage{graphicx,microtype}
\usepackage{amsmath}
\usepackage[bookmarks=false,colorlinks]{hyperref}
\usepackage{lmodern}
\usepackage[all]{hypcap} 
\hypersetup{linkcolor=magenta,citecolor=MidnightBlue,filecolor=Plum,urlcolor=MidnightBlue}
\newcommand{\rev}[1]{\textcolor{black}{#1}}
\newcommand{\revnew}[1]{\textcolor{black}{#1}}



\begin{document}

\title{Effect of Chemical Short-Range Order and Percolation on Passivation in Binary Alloys}

\author{Abhinav Roy}
\affiliation{Department of Materials Science and Engineering, Northwestern University, Evanston, Illinois  60208, USA}
\author{Karl Sieradzki}
\affiliation{Ira A.\ Fulton School of Engineering, Arizona State University, Tempe, Arizona 85287, USA}
\author{James M.\ Rondinelli}
\email{jrondinelli@northwestern.edu}
\affiliation{Department of Materials Science and Engineering, Northwestern University, Evanston, Illinois  60208, USA}
\author{Ian D.\ McCue}
\email{ian.mccue@northwestern.edu}
\affiliation{Department of Materials Science and Engineering, Northwestern University, Evanston, Illinois  60208, USA}

\begin{abstract}
    We develop a percolation model for face centered cubic binary alloys with chemical short-range order (SRO) to account for chemical ordering/clustering that occurs in nominally random solid solutions. We employ a lattice generation scheme that directly utilizes the first nearest neighbor Warren-Cowley SRO parameter to generate the lattice. We quantify the effects of SRO on the first nearest neighbor 3D site percolation threshold using the large cell Monte Carlo renormalization group method and find that the 3D site percolation threshold is a function of the SRO parameter. \revnew{We analyze the effects of SRO on the distribution of the total number of distinct clusters in the percolated structures and find that short-ranged clustering promotes the formation of a dominant spanning cluster.}
    Furthermore, we find that the scaling exponents of percolation are independent of SRO. 
    We also examine the effects of SRO on the 2D-3D percolation crossover and find that the thickness of the thin film for percolation crossover is a function of the SRO parameter. We combine these results to develop a percolation crossover model to understand the electrochemical passivation behavior in binary alloys. The percolation crossover model provides a theoretical framework to understand the critical composition of passivating elements for protective oxide formation. With this model, we show that SRO can be used as a processing parameter to improve corrosion resistance.
\end{abstract}
\date{\today}
\maketitle

\section{Introduction}

Percolation theory has been widely applied to problems spanning biology, epidemics, and complex network theory (see Refs.\  \cite{shante1971introduction,araujo2014recent,li2021percolation} and references therein), making it one of the leading statistical physical models for understanding critical phenomena, e.g., the presence of sharp phase transition points.
It has recently become  instrumental in 
addressing one of the long-standing problems in corrosion science: how does chemical composition and atomic structure dictate primary passivation of an alloy \cite{mccafferty2010introduction}.
In the prototypical binary BCC Fe-Cr system, although Cr is electrochemically more active than Fe, it is kinetically stabilized in the alloy and acts as the passivating element \cite{liu2018effect}. Owing to the selective dissolution of Fe, it has been proposed that the geometric connectivity of Cr clusters is crucial for explaining primary passivation in Fe-Cr alloys through the formation of incipient \{-Cr-O-Cr-\} oxide nuclei \cite{sieradzki1986percolation, qian1990validation,xie2021percolation}. 
These nuclei can be undercut and removed from the surface by the selective Fe dissolution---a process that decreases the Fe surface concentration \cite{newman1988validation}. In the absence of undercutting, we would expect that even for alloys relatively dilute in Cr, selective Fe dissolution would result in the rapid development of a stable passive film. \revnew{For example, consider the behavior of a Fe$_{1-p}$Cr$_{p}$ alloy in which $p$ is the initial atomic fraction of Cr in the alloy. If $h$ monolayers of Fe are selectively dissolved, the atom fraction of Cr in the alloy, $p(h)$, is given by $1-(1-p)^{h}$ \cite{wagner1997dealloying} so that only about $4.3$ layers would be required for the Cr surface concentration to attain a value of about 0.20. This value is equivalent to that of a 300 series stainless steel.} The concept of continuous geometric connectivity of the passivating component leads to the percolation theory of passivation \cite{shante1971introduction, xie2021percolation}. It was proposed that the 3D site percolation threshold (up to the third nearest neighbor) of Cr atoms is central to predicting the critical composition of Cr needed for passive film formation on the alloy \cite{liu2018effect}. 
The dependence of this connectivity on chemical short-range order (SRO) is of keen interest because elemental selection and processing enables SRO control that could be leveraged to improve primary passivation.
Chemical SRO 
in alloys is a crucial factor influencing various properties of metallic alloys. 
It is known to improve the mechanical properties of complex concentrated alloys 
through local fluctuations in stacking fault energies \cite{chen2023chemical}, which impact dislocation motion \cite{taheri2023understanding}. It has been proposed that SRO can affect the corrosion resistance of binary alloys \cite{xie2021percolation}.
Previously, the effects of chemical 
SRO on the variation of percolation threshold in FCC binary alloys have been investigated quantitatively \cite{yu1994correlated} using Warren-Cowley SRO parameters \cite{cowley1960short}, albeit for a limited range of SRO and finite-sized systems. In the context of correlated percolation, the threshold variation due to short-range correlations in binary composites has also been investigated \cite{frary2007correlation}. 

 \revnew{Here, we examine percolation in binary FCC solid solutions using the large cell Monte Carlo renormalization group method (MC-RNG) \cite{reynolds1980large} to obtain accurate estimates of the percolation threshold for FCC alloys containing SRO.} We obtain a relationship between the first nearest neighbor (NN) 3D site percolation threshold and the first NN pair-interaction parameter ($\Delta E$) in a binary alloy, which we identify as the microscopic degree of freedom to tune corrosion resistance and enable novel alloy design through chemical selection and processing parameters for desired SRO.
%
%
We also investigate the effects of SRO on the scaling exponents of percolation to find that they are dimensional invariants. 
We combine this result with the 2D-3D percolation crossover theory \cite{sotta2003crossover, zekri20112d} and find that the thin film thickness is a function of the SRO parameter. 
The percolation crossover model allows us to understand the percolation behavior of thin films, thereby giving insights into the primary electrochemical passivation behavior and providing a theoretical framework to justify using SRO for tuning corrosion resistance.

\section{Methods}
\label{methods}
\subsection{Lattice Generation Scheme}
\label{lattice-scheme}
We consider an FCC A-B alloy to study the effect of SRO on the first NN 3D site percolation threshold of the passivating element (in this case, we consider A atoms to be the passivating component of the alloy). In previous studies, various lattice generation methodologies have been proposed and rigorously examined \cite{fey2022random}. Most of these lattice generation schemes use a Monte Carlo method to induce SRO into the lattice by randomly choosing a pair of atoms in a particular nearest neighbor shell and swapping the atoms with a given acceptance probability. The acceptance probability, using the pair-interaction energy and temperature,  maintains the desired short- and long-range order \cite{gehlen1965computer}. Here, we employ a lattice generation scheme that directly utilizes the Warren-Cowley SRO parameter to calculate the site occupation probability (see Appendix~\ref{sec:pop_scheme} for details of the scheme).

The first NN Warren-Cowley SRO parameter as a function of composition ($\chi$) and temperature 
 ($T$) is defined as \cite{wolverton2000short}:
\begin{equation}
    \alpha^{(1)}(\chi,T) = \alpha = 1 - \frac{p_{\mathrm{AB}}}{\chi_\mathrm{A}} 
\end{equation}
where, \rev{$p_\mathrm{AB}$ is the conditional probability that a B atom will have an A atom as its first NN and  $\chi_\mathrm{A}$ is the composition of A.} Using this information, the site occupation probability of component A ($p = p_\mathrm{A}$) is calculated directly from $\alpha$ using the scheme provided in Appendix~\ref{sec:pop_scheme}. The lattice generation scheme in previous studies is convenient when interaction energies, such as those generated via first principles, can be used as input. 
However, the scheme in Appendix~\ref{sec:pop_scheme} is useful when working directly
with 
extended x-ray absorption fine structure (EXAFS) \cite{fantin2020short} and extended electron energy-loss fine structure (EXELFS) \cite{taheri2023understanding} experimental data, 
where neighbor occupancies are extracted to quantify SRO.
\revnew{We note that the local fluctuations of the SRO parameters do not affect the percolation threshold values as the percolation threshold is a fundamental geometric property of the lattice. Lattices with a given value of SRO parameter have a fixed percolation threshold, as SRO is also an average quantity.}

\begin{figure}
    \centering
    \includegraphics[width=\columnwidth]{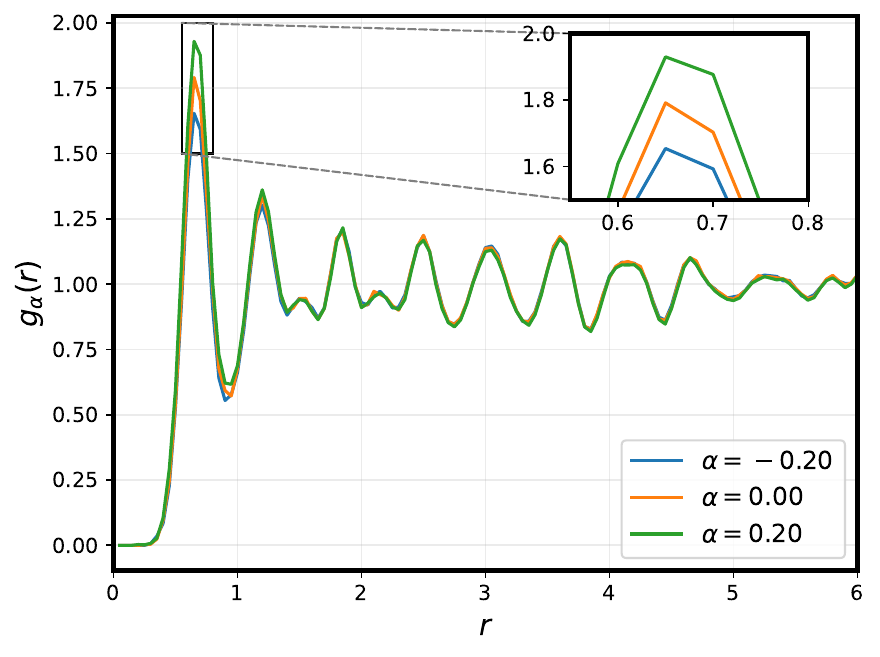}
    \caption{Radial distribution function of component A ($b=32$).}
    \label{Fig1}
\end{figure}

\begin{figure*}
\centering
\includegraphics[width=0.85\textwidth]{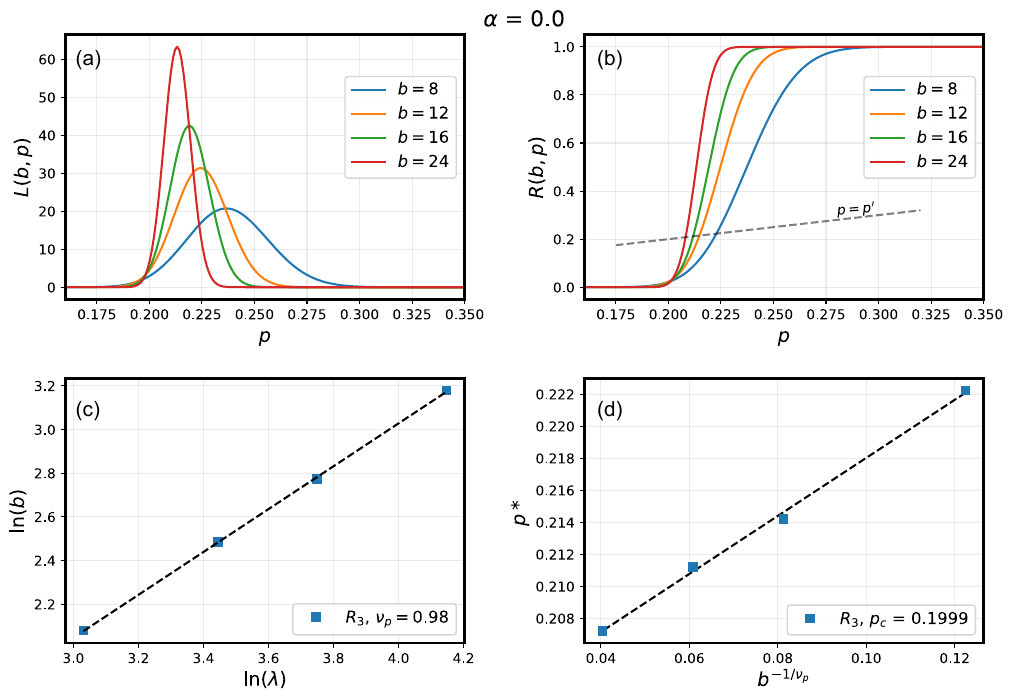}
\caption{Percolation threshold calculated from the MC-RNG analysis for a random binary solid solution on an FCC lattice ($\alpha=0$). (a) \rev{$\beta$ distribution fitted to the histograms of the spanning probabilities for specified cell sizes.} (b) Cumulative distribution, $R(b,p)$, is calculated from the $\beta$ distribution, and the cell percolation threshold, $p^{*}$, is calculated from the point of intersection with line $p=p'$. (c) The natural logarithm of $\lambda_{p}$ versus that of cell size $b$ gives a scaling exponent $\nu_{p} = 0.98$. (d) The cell percolation threshold $p^{*}$ plotted against $b^{-1/\nu_{p}}$ in the limit of $b \rightarrow \infty$.}
\label{Fig3}
\end{figure*}

To verify the lattice generation scheme, we plot in \autoref{Fig1} the radial distribution function (RDF) $g_{\alpha}(r)$ of the passivating element (A) for different values of $\alpha$  using the \texttt{rdfpy} Python module \cite{rdfpy}. The calculated RDF does not include any X-ray or neutron scattering cross-sections. The first peak occurs at a distance $\sim 1/\sqrt{2}$ (the non-dimensionalised lattice constant ($a$) in the model is set to $a=1$).
The RDF is calculated using an FCC lattice cell of size $b = 32$ (i.e., $32$ unit cells per dimension). The first peak for the chemical short-range ordered ($\alpha < 0$) cell is lower than that for the random solid solution ($\alpha = 0$) as shown in the inset of \autoref{Fig1}. 
\rev{For $\alpha>0$ (denotes short-range clustering), the peak increases, which we understand as follows: like atoms will prefer to be closer, giving rise to fewer unlike atom neighbors in the first coordination shell (and vice versa).}

\subsection{Renormalization Group Method}
\label{mcrng}
 We obtain the value of the first NN 3D site percolation threshold,  $p_{c}^{3D} \{1\}$, using a large cell Monte Carlo renormalization group (MC-RNG) method \cite{reynolds1980large}. In percolation theory, the value of $p_{c}^{3D} \{1\}$ is defined for a spanning cluster consisting of A atoms in the infinite lattice. In the MC-RNG scheme, we choose finite cells of various sizes and increase the composition successively until the critical composition ($p$) is obtained, i.e., a spanning cluster occurs. We define the rule ($R_3$) for percolation in those cells to occur if a spanning cluster spans the entire lattice in all three spatial dimensions. $R_3$ is used as it rigorously describes percolation criteria in a finite cell [see the Supplemental Materials (SM) \cite{Supp}]. We note that $R_{3}$ employs open-boundary conditions because treating the cell as a representative element of the infinite lattice is not required in the MC-RNG method \cite{reynolds1980large}. To precisely determine the spanning probability (i.e., the critical composition $p$) in such finite-sized cells, we developed a cluster labeling algorithm for the FCC lattice based on the Hoshen-Kopelman algorithm \cite{hoshen1976percolation, al2003extension}. \rev{A brief description of spanning probability in the current context and how it can be equated to the critical composition $p$ is provided in Section II of SM \cite{Supp}}.
 
 We obtain $p$ from our cluster labeling algorithm for a particular realization and cell size. We perform this calculation for four different cell sizes $b = 8,12,16,24$ and $1000$ different realizations each to obtain the percolation threshold distribution (see Sec.\ S3 of Ref.\ \onlinecite{Supp}). We fit the $\beta$ distribution to the histogram, obtaining the probability density $L(b,p)$ as shown in \autoref{Fig3}a.
 The choice of the $\beta$ distribution to fit the histograms is motivated by characteristics of the $\beta$ function \cite{arfken2011mathematical} given by:
 \begin{equation}
    \beta (u,v) = \int_{0}^{1} p^{u-1} (1 - p)^{v-1} \, dp \quad \forall \quad u,v > 0\,,
\end{equation}
 where we can set the real parameters $u$ and $v$ using the mean $\mu$ and the variance $\sigma^2$ of the percolation threshold distribution. The parameters of the $\beta$ distribution are defined as follows:
$u= \mu \left[\mu (1 - \mu)/\sigma^2 - 1\right]$ and 
$v= (1 - \mu) \left[\mu (1 - \mu)/\sigma^2 - 1\right]$, 
 whereby $\mu$ and $\sigma^{2}$ of the percolation threshold distribution are obtained from the respective histograms in Figure S4 of the SM \cite{Supp}. 
 
The renormalized probability of spanning a cell ($p'$) is defined as the total probability at which a spanning cluster occurs, which is the same as the cumulative distribution function 
$p' = R(b,p) = \int_0^{p} L(b,\hat{p})\, d\hat{p}$.
The probability $R(b,p)$ is also the renormalization rule in MC-RNG. The fixed point of transformation, i.e., the cell percolation threshold ($p^{*}(b)$), is obtained from the point of intersection of $R(b,p)$, and the line $p=p'$ (\autoref{Fig3}b). In percolation theory, the cell percolation threshold scales with the system size according to the expression 
$|p^{*}(b)-p_{c}^{3D} \{1\}| \propto b^{- 1/\nu_{p}},$
where the scaling exponent $\nu_{p}$ is obtained from the renormalized probabilities by the expression 
$\nu_{p} = \ln(b)/\ln(\lambda_p),$ 
as shown in \autoref{Fig3}c. Here,
\begin{equation}
\left. \lambda_p = \frac{dR(b,p)}{dp} \right|_{p=p'}.
\end{equation}
Plotting $b^{-1/\nu_{p}}$ against the cell percolation threshold gives $p_{c}^{3D} \{1\} = 0.1999$ for the random FCC lattice as $b \rightarrow \infty$ (\autoref{Fig3}d), which is within $0.35\%$ of the accepted value of $0.1992$ \cite{lorenz2000similarity}.
\section{Results and Discussion}
\label{results}
\subsection{Percolation Threshold Variation with SRO}
\label{pc}
 The first NN pair-interaction parameter $\Delta E = E_{AB} - (E_{AA} + E_{BB})/2$ (units of meV per atom), whereby 
 $E_{AA}$, $E_{AB}$, and $E_{AB}$ represent the bond energy (assigned a negative value) of A-A, B-B, and A-B atoms, respectively,  
 is related to $\alpha$ as \cite{cheng1967vacancy}: 
\begin{equation}
    \frac{(1-\alpha)^2}{\left(\frac{\chi_\mathrm{A}}{\chi_\mathrm{B}} + \alpha \right) \left(\frac{\chi_\mathrm{B}}{\chi_\mathrm{A}} + \alpha \right)} = \exp\left(-\frac{z\Delta E}{k_{B}T}\right)
    \label{eqn8}
\end{equation}
where $z$ denotes the coordination number of the lattice (in the case of FCC, $z=12$), $\chi_\mathrm{A}$ and $\chi_\mathrm{B}$ represent the mole fractions of components A and B, $k_{B}$ denotes the Boltzmann constant and $T$ represents the temperature (in Kelvin) of the system which is set to $300$\,K. 

\begin{figure}
\centering
\includegraphics[width=\columnwidth]{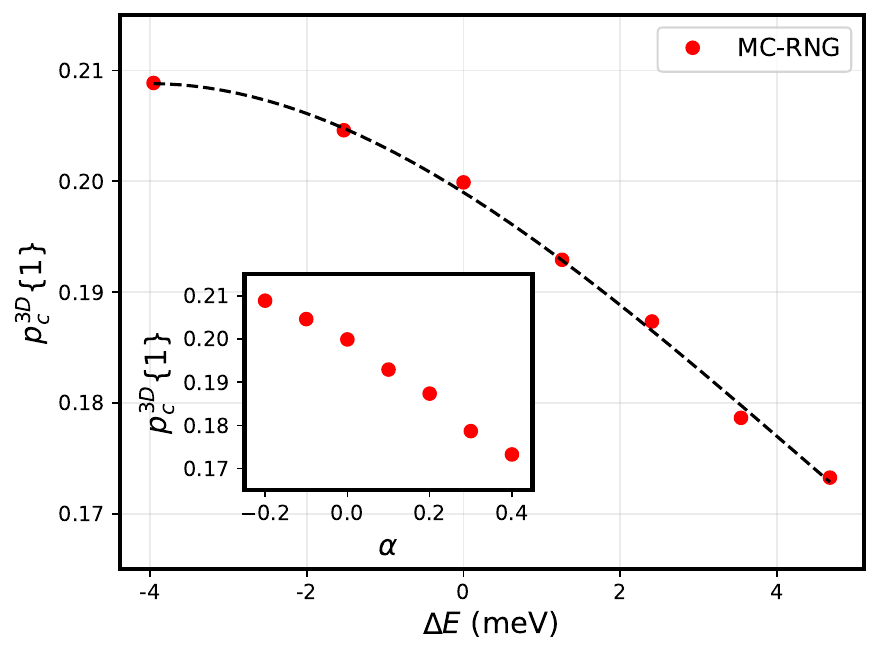}
\caption{The variation of first NN 3D site percolation threshold with the first NN pair-interaction parameter $\Delta E$. The inset shows the same variation with SRO parameter $\alpha$. \revnew{The percolation thresholds are obtained from the MC-RNG analysis, and the dashed line is a guide to the eye.}}
\label{Fig4}
\end{figure}

\autoref{Fig4} shows the dependence of $p_{c}^{3D} \{1\}$ versus both the first NN pair-interaction parameter ($-5 \leq \Delta E \leq 5\,\mathrm{meV}$) and Warren-Cowley SRO parameter ($-0.2 \leq \alpha \leq 0.4$). 
\rev{When relating the $\Delta E$  and $\alpha$ parameters to plot \autoref{Fig4}, we used the percolation threshold calculated from the MC-RNG method for the composition variable}.
The percolation threshold decreases with increasing values of $\Delta E$, which indicates that short-ranged clustering promotes the lowering of $p_{c}^{3D} \{1\}$. This observation is consistent with the behavior of $p_{c}^{3D}\{1\}$ in BCC alloys with SRO \cite{xie2021percolation}. \rev{A comparison of $p_{c}^{3D}\{1\}$ variation with $\Delta E$ for the FCC lattice (this study) and for the BCC lattice (Ref.\  \onlinecite{xie2021percolation}) is presented in Fig.\ S5 of the SM \cite{Supp}.}

We fit a third degree polynomial to the data in \autoref{Fig4} given by $p_{c}^{3D}\{1\} = p_{c,0}^{3D} + \theta_{1}\Delta E + \theta_{2}\Delta E^{2} + \theta_{3}\Delta E^{3}$, 
where $p_{c,0}^{3D}$ represents the percolation threshold for a random alloy ($\alpha=0$). The confidence-of-fit is $R^2=0.997$ 
for the analytical model 
with $\theta_{1} = -4.437$, $\theta_{2} = -3.808 \times 10^{2}$, $\theta_{3} = 2.872 \times 10^{4}$, respectively.  We find that the relative standard deviation of percolation threshold values for the given range of $\Delta E$ parameter is about $6.39 \%$. \rev{The polynomial fit enables the interpolation of values for the percolation threshold over the range of $\Delta E \in [-5,5] \, \mathrm{meV}$ for an FCC lattice without performing the MC-RNG calculations}. This analysis may be extended up to the third nearest neighbor to calculate the variation of the corresponding site percolation threshold $p_{c}^{3D}\{1,2,3\}$ with SRO.

\subsection{\revnew{Cluster-Quantity Fluctuation}}
\label{cluster-number}
\revnew{The total number of distinct clusters of the passivating component in the alloy at the percolation threshold is crucial to understanding the effects of SRO on percolation. The variation in the number of clusters, referred to as cluster quantity fluctuation, can be quantified by examining the distribution of the number of clusters. We specifically focus on the variation of the number of distinct clusters found at the percolation threshold, including the spanning cluster, for a range of cell sizes.}

\begin{figure*}
\centering
\includegraphics[width=0.85\textwidth]{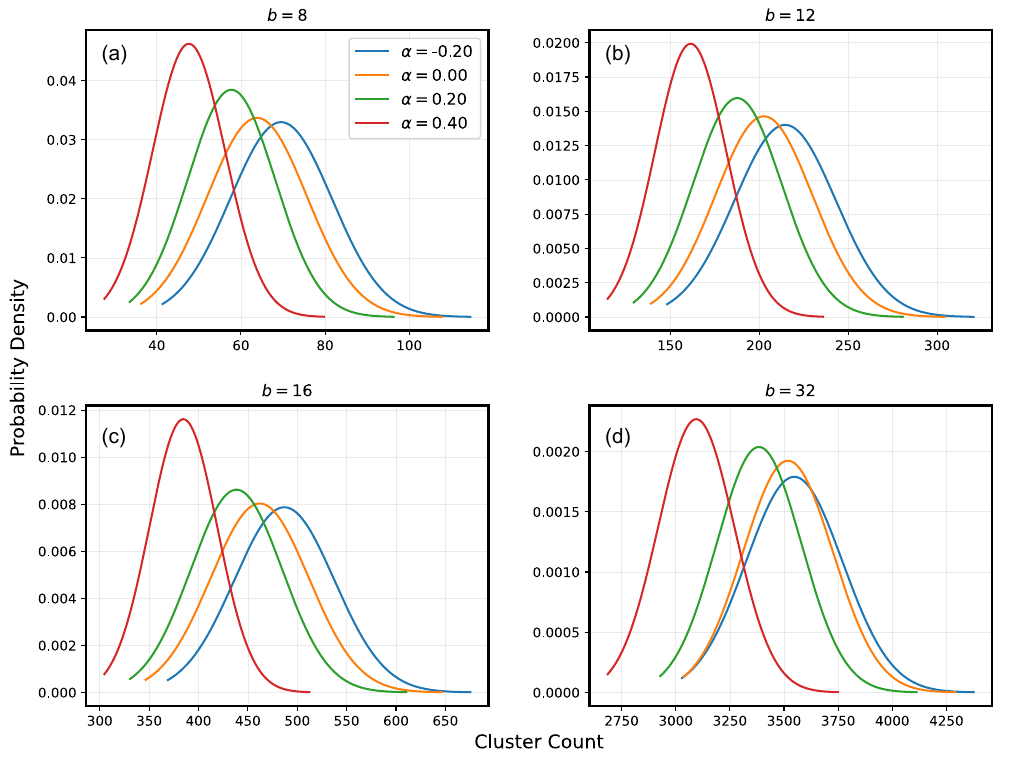}
\caption{\revnew{Cluster-quantity fluctuation} is described as the Gaussian distribution of the number of distinct clusters at the percolation threshold of a particular cell size for 1000 realizations. The fluctuation is shown for (a) $b=8$ (b) $b=12$ (c) $b=16$ (d) $b=32$.}
\label{Fig5}
\end{figure*}

Therefore, we calculate the total number of clusters in a particular cell of size $b=8,12,16,32$ (1000 realizations each) at the \rev{cell percolation threshold $p^{*}(b)$}. We then calculate the probability density function for different cell sizes (\autoref{Fig5}), to which we fit a Gaussian distribution.  
We find that for higher values of $\alpha$ the distribution of cluster number shifts toward the left, signifying a lower number of distinct clusters at the percolation threshold. Clustering promotes the formation of a dominant spanning cluster thereby reducing the total number of distinct clusters found at the percolation threshold. \rev{This also explains the lowering of $p_{c}^{3D} \{1\}$  with increasing value of $\Delta E$, as positive values of $\Delta E$ means short-range clustering in the system.} 
\subsection{Scaling Exponents}
\label{critical-exponents}
It is known that the scaling exponents of percolation are universal, i.e., they are independent of the lattice type for a given dimension \cite{shante1971introduction}. The scaling exponent $\nu_{p}$ corresponds to the exponent related to the correlation length ($\xi$) \cite{li2021percolation} of percolation. For long-range correlated percolation, where the spatial correlations follow a power law behavior given by $C(r) \sim r^{-m}$ \cite{zierenberg2017percolation}, the scaling exponent is modified according to the degree of correlation ($m$) \cite{weinrib1984long}. For short-ranged correlations the scaling exponent is invariant; it is the same as that of an uncorrelated (random) system.   
\begin{figure}
\centering
\includegraphics[width=\columnwidth]{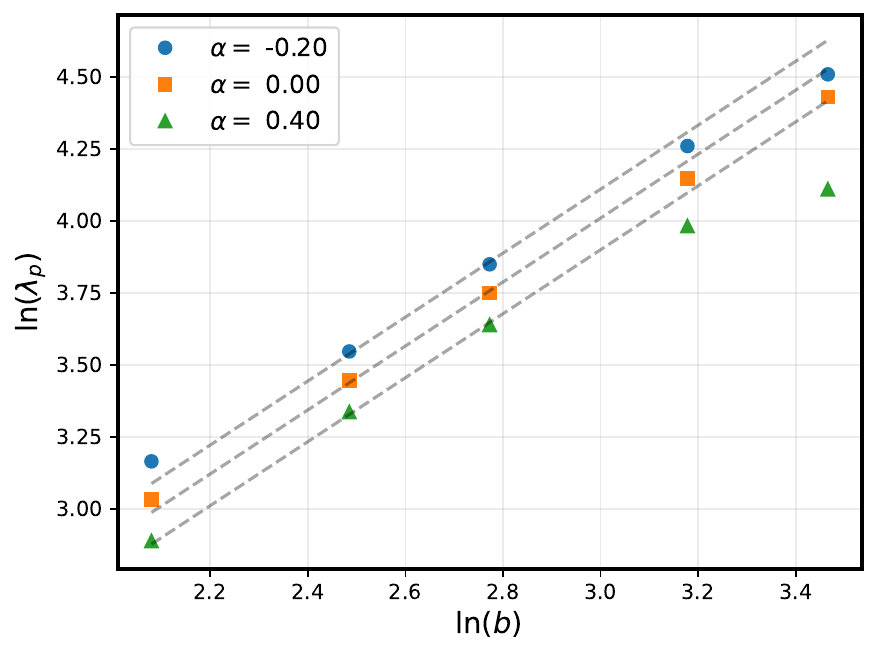}
\caption{The slope of $\ln(\lambda_{p})$ vs $\ln(b)$ gives $1/\nu_{p}$ asymptotically.}
\label{Fig6}
\end{figure}

We confirm that the true value of scaling exponent $\nu_{p}$ is independent of SRO in the FCC structure (\autoref{Fig6}). From the renormalization group method, the true value of the scaling exponent $\nu_{p}$ is obtained from the inverse of the slope of $\ln(\lambda_{p})$ plotted against $\ln(b)$ as $\lambda_{p} = b^{1/\nu_{p}}$ \cite{reynolds1980large}. In \autoref{Fig6}, the straight dashed lines represent lines with slope $\nu_{p}^{-1}$, which matches with the data asymptotically. From the straight line fit, we validate the true value of $\nu_{p} = \nu_{3D} \approx 0.875$ \cite{lorenz1998precise} for different values of $\alpha$. The value of $\nu_{p}$ calculated from MC-RNG is crucial in the model for 2D-3D percolation crossover in FCC alloys with SRO.

 We also report the calculated values of the scaling exponent $\tau$ related to the normalized cluster size distribution denoted by  $n(s)$, which represents the number of clusters of size $s$ per lattice site. For large clusters $s$ at the percolation threshold, $n(s) \sim s^{-\tau}$ \cite{ding2014numerical}. We calculate this distribution for different values of $\alpha$ (see the SM \cite{Supp}) and show that $\tau$ is also invariant for a given dimension.

\subsection{Percolation Crossover Model}
\label{crossover}
Owing to the selective dissolution of the non-passivating element that precedes the primary passivation process, percolation across a 2D roughened surface occurs below the corresponding 2D percolation threshold \cite{xie2021percolation}. The critical composition for passivation has a lower bound set by the 3D site percolation threshold, and it is related to the number of monolayers of selective dissolution. Therefore, the phenomena of a 2D-3D percolation crossover effect \cite{zekri20112d} is crucial for explaining the value of this critical passivating element composition. \rev{Such a crossover effect describes the percolation phenomena in thin films where the percolation behavior is intermediate between a 2D and a 3D system. In this transition regime between pure 2D percolation behavior and pure 3D percolation behavior, we find the percolation threshold of the thin film is only dependent on the thickness of the film.} To that end, understanding passivation in metallic alloys where the passive film is essentially thin is necessary. A percolation model for such crossover effects was proposed by Sotta and Long \cite{sotta2003crossover} for a random structure. 

Here, we show that such a model is sensitive to SRO in the lattice. The 2D-3D percolation crossover model which relates the film thickness $h$ as a function of $\alpha$ to the thickness-dependent percolation threshold, $p_{c}(h,\alpha)$ (also a function of $\alpha$), is given as 
\begin{equation}
    h(\alpha) = \delta[p_{c}(h,\alpha) - p_{c}^{3D}(\alpha)]^{-\nu_{p}}
    \label{eq11}
\end{equation}
where $\delta$ is a fitting parameter and $p_{c}^{3D}(\alpha)$ is the bulk 3D site percolation threshold of an FCC lattice (for a given $\alpha$). In the percolation crossover model, the value of $\Delta E$ sets the value of $\alpha$ at a specific composition and temperature, hence for that composition and temperature (by setting $p_{c}(h,\alpha)$ and $T$), $h$ becomes a function of $\alpha$. The true value of $\nu_{p} = \nu_{3D}$ is used in the model described in Ref.\  \onlinecite{sotta2003crossover} for a random structure. In our analysis, we use the value of $\nu_{p}$ calculated from the renormalization group method to account for the finite size effects. 
\begin{figure}
\centering
\includegraphics[width=\columnwidth]{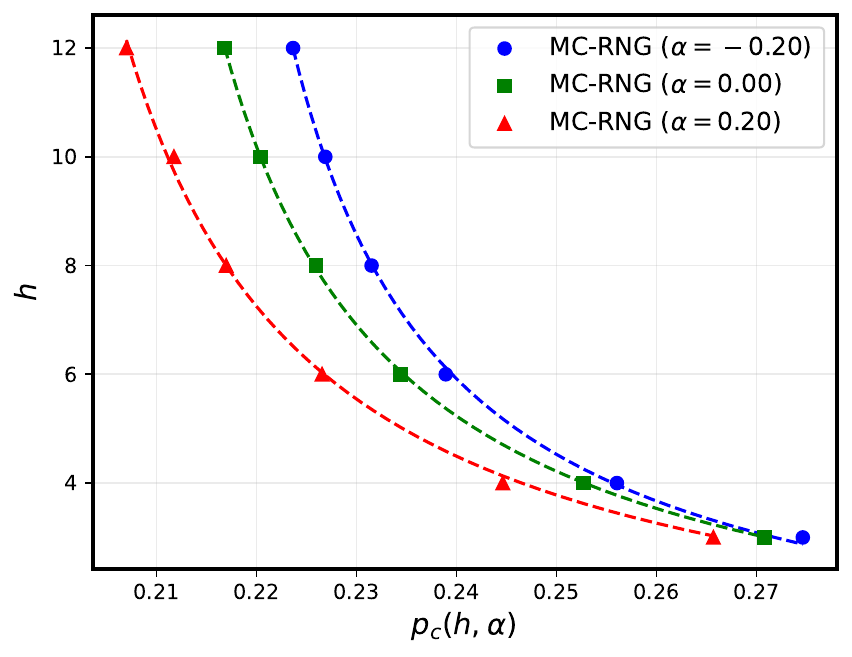}
\caption{\revnew{The calculated MC-RNG values (data) and fit to 2D-3D percolation crossover model (\autoref{eq11}) for different SRO parameters  $\alpha$.}}
\label{Fig7}
\end{figure}

 \autoref{Fig7} plots $p_{c}(h,\alpha)$ for various thin film layer thicknesses ($h$). \rev{In our calculations, $h$ is the thickness of the thin film where the composition is such that it matches the percolation threshold of the thin film.} Each data point for different $\alpha$ corresponds to the percolation threshold calculated using MC-RNG (Sec. \ref{mcrng}). Owing to significant finite-size effects in the 2D to 3D percolation transition regime, we account for finite size correction to the scaling exponent in our MC-RNG analysis \cite{nobrega2003corrections}. The modified scaling relation for the cell percolation threshold becomes $p^{*}(b) \propto b^{- 1/\nu_{p} - \omega}$, where the correction to scaling exponent is $\omega \approx 0.9$ \cite{ziff2002convergence}. In the MC-RNG analysis, we consider cells of dimension $b \times b \times h$ and for a fixed $h$ we get the values of $p^{*}(b)$ for $b=16,24,32,48,64$ from $500$ distinct realizations each. 
 \rev{We have maintained uniformity in our MC-RNG analysis by using $R_{3}$ for the calculation of $p^{*}(b)$ in this case}. 
 Using the modified scaling relation, we calculate $p_{c}(h,\alpha)$ for a given $h$ and perform this analysis for different values of $h = 3, 4, 6, 8, 10, 12$, as shown in \autoref{Fig7}. We fit the data to the percolation crossover model in \autoref{eq11} and report the values of fitting parameters $\delta$ and $p_{c}^{3D}(\alpha) = p_{c}^{3D}\{1\}$ in table \autoref{tab2}. We also report the calculated values of $p_{c}^{3D}\{1\}$ from \autoref{Fig4} to compare with fitted $p_{c}^{3D}\{1\}$ values. The agreement between the fitted and calculated values of $p_{c}^{3D}\{1\}$ indicates that the percolation crossover model captures the effects of SRO on percolation threshold variation and can be used as a tool to analyze alloys with SRO. 
 
 A recent study of the FCC Cu-Rh alloy \cite{xie2021passivation} showed that the passivation behavior was in quantitative agreement with the percolation crossover model. From the fitting parameter obtained (third NN percolation threshold $p_{c}^{3D}\{1,2,3\}$), the authors qualitatively predicted that the alloy shows a tendency towards short-ranged ordering; however, the experimental verification of this claim is pending. 
 \rev{We show in \autoref{Fig6} that our model given by \autoref{eq11} captures the variation in percolation threshold $p_{c}^{3D}\{1\}$ (extracted from the fitting parameters). If the same analysis is carried out for $p_{c}^{3D}\{1,2,3\}$, a \emph{quantitative} prediction about SRO in the alloy can be obtained. This would be done by fitting their experimental data to our model.} Such experimental verification will also validate the model proposed in this study, which can be matched quantitatively with the data obtained from the MC-RNG calculations. 
 
\begin{table}
\centering
\caption{Values of the fitting parameters $\delta$ and $p_{c}^{3D}\{1\}$ from the percolation crossover model along with the calculated values of $p_{c}^{3D}\{1\}$ extracted from \autoref{Fig4}.}
\begin{ruledtabular}
\begin{tabular}{cccc}
 $\alpha$ & $\delta$ &  $p_{c}^{3D}\{1\}$ (fitted)  &  $p_{c}^{3D}\{1\}$ (calculated) \\ 
 \hline
$-0.2 $   &   $0.2017$  &   $0.2082$    &   $0.2088$ \\
$0.0$ &  $ 0.2262$  & $ 0.1994$ & $0.1999$ \\
$0.2 $&   $0.2468$  &   $0.1882$  &  $ 0.1873$  \\ 
\end{tabular}
\end{ruledtabular}
\label{tab2}
\end{table}

We note another observation from \autoref{Fig7}. For a given value of thin film composition $p_{c}(h,\alpha)$, the value of $h$ for an alloy exhibiting short-ranged clustering ($\alpha > 0$) is lower than the random solid solution ($\alpha = 0$), which in turn is lower than that value of an alloy with short-ranged order ($\alpha < 0$). Because the value of $h$ corresponds to the number of monolayers dissolved due to the selective dissolution process, \rev{the criterion for better corrosion resistance is to have a thinner film thickness ($h$) value.} Therefore, short-ranged clustering leads to better corrosion resistance than short-range ordering.
\section{Conclusion}
\label{conclusion}
In this work, we developed a percolation model of the 2D-3D percolation crossover phenomenon for alloys with SRO to study the electrochemical passivation behavior. This model can be used to understand how to tune SRO to modify the corrosion resistance of alloys through novel processing parameter spaces - to lock in preferential order -- or through chemical selection that controls $\Delta E$. 
Our main results and conclusions include: 
\begin{enumerate}
    \item Use of a lattice generation scheme that can induce desired SRO, which we 
    validated by calculating the RDF of the passivating component. We show the effects of SRO on the first NN peak value and report that the peak value increases with short-range clustering.
    \item Demonstration of the MC-RNG method to calculate the first NN 3D site percolation thresholds. We find that the percolation threshold is a function of SRO and propose an analytical expression relating the threshold with $\Delta E$. 
    \item Formulation of a percolation crossover model that accounts for SRO. This model also suggests that short-ranged clustering promotes corrosion resistance in binary alloys.
\end{enumerate}

\rev{The current study for incorporating SRO effects utilizes ideal conditions (for example, the absence of defects or voids), which can then be subsequently expanded to include further dependencies found in real alloys. To that end, our motivation for this study was to establish a theoretical framework for understanding the electrochemical passivation behavior using model binary alloys.}
Chemical SRO in alloys is characterized using various experimental techniques such as diffuse x-ray scattering, transmission electron microscopy (TEM), EXAFS, and EXELFS \cite{fantin2020short, taheri2023understanding, walsh2023reconsidering}. Recently, SRO in complex concentrated alloys was re-evaluated and it was suggested that existing experimental techniques might be inadequate for the quantitative characterization of SRO due to intricate interactions among different constituents \cite{walsh2023reconsidering}. We propose that the percolation crossover model serves as a tool that can be used to validate the presence of SRO in alloys using electrochemical experiments. When this model is coupled with the proposed polynomial fit for $p_{c}^{3D}\{1\}$, it can give a quantitative estimate of $\Delta E$ for the alloy system. This analysis can then be used to calibrate and validate first-principles simulations of binary alloys, which typically model chemical SRO using cluster expansions \cite{kadkhodaei2021cluster}.

 The effects of SRO on percolation threshold in FCC alloys also serve as a foundation for further research where effects of SRO can be investigated in other domains, for example, in complex networks \cite{berchenko2009emergence} or networks of neurons \cite{tian2022percolation}. Recently, the phenomenon of 1D wormhole corrosion in Ni-Cr alloys was discovered \cite{yang2023one}, where accelerated localized corrosion occurs due to seepage of corrosive salt solution through 1D percolating voids in the grain boundaries. It would be interesting to study the effects of SRO in such a system and how it influences the percolation of voids in 1D wormhole corrosion. 
\begin{acknowledgments}
This work was supported by the National Science Foundation (NSF) under award numbers DMR-2208865 (A.R., J.M.R., and I.D.M.) and DMR-2208848 (K.S.). We thank Aden Weiser for his contributions to the initial version of lattice generation code. This research was supported in part through the computational resources and staff contributions provided for the Quest high performance computing facility at Northwestern University which is jointly supported by the Office of the Provost, the Office for Research, and Northwestern University Information Technology. This work used Bridges-2 at Pittsburgh Supercomputing Center through allocation mat230008 from the Advanced Cyberinfrastructure Coordination Ecosystem: Services \& Support (ACCESS) program, which is supported by National Science Foundation grants \#2138259, \#2138286, \#2138307, \#2137603, and \#2138296.
\end{acknowledgments}
\appendix
\renewcommand{\thefigure}{A\arabic{figure}}
\renewcommand{\thetable}{A\arabic{table}}
\setcounter{figure}{0}
\setcounter{table}{0}

\section{\label{sec:pop_scheme}Site Occupation Probability}
We describe the procedure for calculating the site occupation probability ($p_\mathrm{A}$) central to the lattice generation scheme. We adopt the scheme from Ref. \cite{xie2021passivation} and modify it for an FCC lattice. In Sec.\  \ref{lattice-scheme} we defined the probability $p_\mathrm{AB}$. We define the probabilities $p_\mathrm{BA}$, $p_\mathrm{AA}$ and $p_\mathrm{BB}$ in terms of $\alpha$ as follows:
\begin{align}
    p_{AB} &= \chi_\mathrm{A}(1-\alpha) \\
    p_{BA} &= \chi_\mathrm{B}(1-\alpha) \\
    p_{AA} &= \chi_\mathrm{A} + \alpha\chi_\mathrm{B} \\
    p_{BB} &= \chi_\mathrm{B} + \alpha \chi_\mathrm{A}
\end{align}
where $\chi_\mathrm{A}$ and $\chi_\mathrm{B}$ denote the global atom fraction of components A and B, respectively. We start with an empty FCC cell of a particular size $b$. Thereafter, we populate the cell with a total of $4b^3$ sites by randomly picking an empty lattice site in the cell and populating it with A atoms based on $p_\mathrm{A}$. The site occupation probability is calculated as follows:
\begin{align}
    p''_\mathrm{A} &= N_\mathrm{A}p_\mathrm{AA} + N_\mathrm{B}p_\mathrm{AB} \\
    p''_\mathrm{B} &= N_\mathrm{B}p_\mathrm{BB} + N_\mathrm{A}p_\mathrm{BA} \\
    p_\mathrm{A} &= \frac{p''_\mathrm{A}}{p''_\mathrm{A} + p''_\mathrm{B}} \\
    p_\mathrm{B} &= \frac{p''_\mathrm{B}}{p''_\mathrm{A} + p''_\mathrm{B}}
\end{align}
where $N_\mathrm{A}$ and $N_\mathrm{B}$ correspond to the number of A and B atoms in the first NN shell \rev{of the randomly picked site}, respectively. For given values of $N_\mathrm{A}$, $N_\mathrm{B}$, and $\alpha$, we can calculate $p_\mathrm{A}$. \rev{Throughout the process of populating empty sites, $0 \leq N_{A} + N_{B} \leq 12$.}

\rev{The lattice generation scheme produces single “snapshot” atomic configurations. We check the final global composition of the generated lattice with the initial global composition ($\chi_\mathrm{A}$, $\chi_\mathrm{B}$) to ensure that the number of A-A, B-B, and A-B bonds does not deviate from the desired value for a given $\alpha$. At the initialization of the procedure, $N_\mathrm{A} = N_\mathrm{B} = 0$. In this case, the site occupation probability is simply the global composition of the passivating component, i.e., $p_\mathrm{A} = \chi_\mathrm{A}$. Only when either of $N_\mathrm{A}$ or $N_\mathrm{B}$ or both becomes non-zero, the procedure is employed.} \rev{We would like to note that the empty sites present during the lattice generation scheme are distinct from the empty sites that may occur due to the selective dissolution process. The final lattice generated by our scheme is fully occupied without any empty sites.}

\begin{figure}
\centering
\includegraphics[width=0.95\columnwidth]{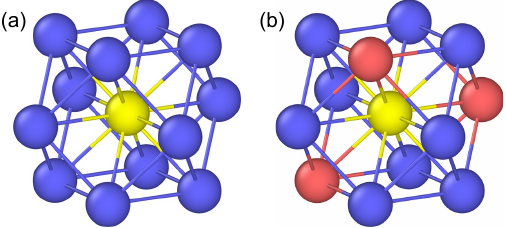}
\caption{(a) A random empty site (yellow atom) in an FCC lattice with 12 B atoms (blue atoms) in the first NN shell. (b) Random empty site with 3 A (red) atoms and 9 B atoms in the first NN shell.}
\label{Fig8}
\end{figure}

We illustrate this scheme with two cases as shown in \autoref{Fig8}. In \autoref{Fig8}a, we find that the random empty site is denoted by a yellow-colored central atom. All the sites of its first NN shell are occupied by the  B atoms (represented by blue-colored atoms). In this case $N_\mathrm{A} = 0$, $N_\mathrm{B} = 12$ and $p''_\mathrm{A} = 12p_\mathrm{AB} = 12\chi_\mathrm{A}(1-\alpha)$. Also,  $p''_\mathrm{B} = 12p_\mathrm{BB} = 12\chi_{B} + 12\alpha\chi_\mathrm{A}$. Therefore, 
\rev{
\begin{equation}
    p_{A} = \chi_\mathrm{A}(1-\alpha)\,.
\end{equation}
}

In \autoref{Fig8}b, we find that the first NN shell of the central random site is filled with 3 A atoms (represented by red-colored atoms) and 9 B atoms. In this case, $N_\mathrm{A} = 3$, $N_\mathrm{B} = 9$ and $p''_\mathrm{A} = 3p_\mathrm{AA} + 9p_\mathrm{AB} = 12\chi_\mathrm{A} + 3\alpha(\chi_\mathrm{B} - 3\chi_\mathrm{A})$. Also, $p''_\mathrm{B} = 9p_\mathrm{BB} + 3p_\mathrm{BA} = 12\chi_\mathrm{B} + 3\alpha(3\chi_\mathrm{A} - \chi_\mathrm{B})$. Therefore, 
\rev{
\begin{equation}
    p_\mathrm{A} = \chi_\mathrm{A} + \frac{\alpha}{4}(\chi_\mathrm{B} - 3\chi_\mathrm{A})\,.
\end{equation}
}
\rev{The site occupation probability ($p_\mathrm{A}$) calculated with the described procedure is used to populate the empty site, which is randomly selected. This is done by generating a random number ($u \in [0,1]$) from a uniform distribution. The random number ($u$) is compared with the calculated site occupation probability $p_\mathrm{A}$. If $u <p_\mathrm{A}$, the randomly selected empty site is populated with atom A, otherwise it is populated with atom B}. We iterate this scheme over all the lattice sites to generate a cell with the desired SRO.

\bibliography{refs_paper1}

\begin{thebibliography}{41}%
\makeatletter
\providecommand \@ifxundefined [1]{%
 \@ifx{#1\undefined}
}%
\providecommand \@ifnum [1]{%
 \ifnum #1\expandafter \@firstoftwo
 \else \expandafter \@secondoftwo
 \fi
}%
\providecommand \@ifx [1]{%
 \ifx #1\expandafter \@firstoftwo
 \else \expandafter \@secondoftwo
 \fi
}%
\providecommand \natexlab [1]{#1}%
\providecommand \enquote  [1]{``#1''}%
\providecommand \bibnamefont  [1]{#1}%
\providecommand \bibfnamefont [1]{#1}%
\providecommand \citenamefont [1]{#1}%
\providecommand \href@noop [0]{\@secondoftwo}%
\providecommand \href [0]{\begingroup \@sanitize@url \@href}%
\providecommand \@href[1]{\@@startlink{#1}\@@href}%
\providecommand \@@href[1]{\endgroup#1\@@endlink}%
\providecommand \@sanitize@url [0]{\catcode `\\12\catcode `\$12\catcode
  `\&12\catcode `\#12\catcode `\^12\catcode `\_12\catcode `\%12\relax}%
\providecommand \@@startlink[1]{}%
\providecommand \@@endlink[0]{}%
\providecommand \url  [0]{\begingroup\@sanitize@url \@url }%
\providecommand \@url [1]{\endgroup\@href {#1}{\urlprefix }}%
\providecommand \urlprefix  [0]{URL }%
\providecommand \Eprint [0]{\href }%
\providecommand \doibase [0]{https://doi.org/}%
\providecommand \selectlanguage [0]{\@gobble}%
\providecommand \bibinfo  [0]{\@secondoftwo}%
\providecommand \bibfield  [0]{\@secondoftwo}%
\providecommand \translation [1]{[#1]}%
\providecommand \BibitemOpen [0]{}%
\providecommand \bibitemStop [0]{}%
\providecommand \bibitemNoStop [0]{.\EOS\space}%
\providecommand \EOS [0]{\spacefactor3000\relax}%
\providecommand \BibitemShut  [1]{\csname bibitem#1\endcsname}%
\let\auto@bib@innerbib\@empty
\bibitem [{\citenamefont {Shante}\ and\ \citenamefont
  {Kirkpatrick}(1971)}]{shante1971introduction}%
  \BibitemOpen
  \bibfield  {author} {\bibinfo {author} {\bibfnamefont {V.~K.}\ \bibnamefont
  {Shante}}\ and\ \bibinfo {author} {\bibfnamefont {S.}~\bibnamefont
  {Kirkpatrick}},\ }\bibfield  {title} {\bibinfo {title} {An introduction to
  percolation theory},\ }\href@noop {} {\bibfield  {journal} {\bibinfo
  {journal} {Advances in Physics}\ }\textbf {\bibinfo {volume} {20}},\ \bibinfo
  {pages} {325} (\bibinfo {year} {1971})}\BibitemShut {NoStop}%
\bibitem [{\citenamefont {Ara{\'u}jo}\ \emph {et~al.}(2014)\citenamefont
  {Ara{\'u}jo}, \citenamefont {Grassberger}, \citenamefont {Kahng},
  \citenamefont {Schrenk},\ and\ \citenamefont {Ziff}}]{araujo2014recent}%
  \BibitemOpen
  \bibfield  {author} {\bibinfo {author} {\bibfnamefont {N.}~\bibnamefont
  {Ara{\'u}jo}}, \bibinfo {author} {\bibfnamefont {P.}~\bibnamefont
  {Grassberger}}, \bibinfo {author} {\bibfnamefont {B.}~\bibnamefont {Kahng}},
  \bibinfo {author} {\bibfnamefont {K.}~\bibnamefont {Schrenk}},\ and\ \bibinfo
  {author} {\bibfnamefont {R.~M.}\ \bibnamefont {Ziff}},\ }\bibfield  {title}
  {\bibinfo {title} {Recent advances and open challenges in percolation},\
  }\href@noop {} {\bibfield  {journal} {\bibinfo  {journal} {The European
  Physical Journal Special Topics}\ }\textbf {\bibinfo {volume} {223}},\
  \bibinfo {pages} {2307} (\bibinfo {year} {2014})}\BibitemShut {NoStop}%
\bibitem [{\citenamefont {Li}\ \emph {et~al.}(2021)\citenamefont {Li},
  \citenamefont {Liu}, \citenamefont {L{\"u}}, \citenamefont {Hu},
  \citenamefont {Xu},\ and\ \citenamefont {Zhang}}]{li2021percolation}%
  \BibitemOpen
  \bibfield  {author} {\bibinfo {author} {\bibfnamefont {M.}~\bibnamefont
  {Li}}, \bibinfo {author} {\bibfnamefont {R.-R.}\ \bibnamefont {Liu}},
  \bibinfo {author} {\bibfnamefont {L.}~\bibnamefont {L{\"u}}}, \bibinfo
  {author} {\bibfnamefont {M.-B.}\ \bibnamefont {Hu}}, \bibinfo {author}
  {\bibfnamefont {S.}~\bibnamefont {Xu}},\ and\ \bibinfo {author}
  {\bibfnamefont {Y.-C.}\ \bibnamefont {Zhang}},\ }\bibfield  {title} {\bibinfo
  {title} {Percolation on complex networks: {T}heory and application},\
  }\href@noop {} {\bibfield  {journal} {\bibinfo  {journal} {Physics Reports}\
  }\textbf {\bibinfo {volume} {907}},\ \bibinfo {pages} {1} (\bibinfo {year}
  {2021})}\BibitemShut {NoStop}%
\bibitem [{\citenamefont {McCafferty}(2010)}]{mccafferty2010introduction}%
  \BibitemOpen
  \bibfield  {author} {\bibinfo {author} {\bibfnamefont {E.}~\bibnamefont
  {McCafferty}},\ }\href@noop {} {\emph {\bibinfo {title} {Introduction to
  corrosion science}}}\ (\bibinfo  {publisher} {Springer Science \& Business
  Media},\ \bibinfo {year} {2010})\BibitemShut {NoStop}%
\bibitem [{\citenamefont {Liu}\ \emph {et~al.}(2018)\citenamefont {Liu},
  \citenamefont {Aiello}, \citenamefont {Xie},\ and\ \citenamefont
  {Sieradzki}}]{liu2018effect}%
  \BibitemOpen
  \bibfield  {author} {\bibinfo {author} {\bibfnamefont {M.}~\bibnamefont
  {Liu}}, \bibinfo {author} {\bibfnamefont {A.}~\bibnamefont {Aiello}},
  \bibinfo {author} {\bibfnamefont {Y.}~\bibnamefont {Xie}},\ and\ \bibinfo
  {author} {\bibfnamefont {K.}~\bibnamefont {Sieradzki}},\ }\bibfield  {title}
  {\bibinfo {title} {The effect of short-range order on passivation of
  {F}e-{C}r alloys},\ }\href@noop {} {\bibfield  {journal} {\bibinfo  {journal}
  {Journal of The Electrochemical Society}\ }\textbf {\bibinfo {volume}
  {165}},\ \bibinfo {pages} {C830} (\bibinfo {year} {2018})}\BibitemShut
  {NoStop}%
\bibitem [{\citenamefont {Sieradzki}\ and\ \citenamefont
  {Newman}(1986)}]{sieradzki1986percolation}%
  \BibitemOpen
  \bibfield  {author} {\bibinfo {author} {\bibfnamefont {K.}~\bibnamefont
  {Sieradzki}}\ and\ \bibinfo {author} {\bibfnamefont {R.}~\bibnamefont
  {Newman}},\ }\bibfield  {title} {\bibinfo {title} {A percolation model for
  passivation in stainless steels},\ }\href@noop {} {\bibfield  {journal}
  {\bibinfo  {journal} {J. Electrochem. Soc}\ }\textbf {\bibinfo {volume}
  {133}},\ \bibinfo {pages} {1979} (\bibinfo {year} {1986})}\BibitemShut
  {NoStop}%
\bibitem [{\citenamefont {Qian}\ \emph {et~al.}(1990)\citenamefont {Qian},
  \citenamefont {Newman}, \citenamefont {Cottis},\ and\ \citenamefont
  {Sieradzki}}]{qian1990validation}%
  \BibitemOpen
  \bibfield  {author} {\bibinfo {author} {\bibfnamefont {S.}~\bibnamefont
  {Qian}}, \bibinfo {author} {\bibfnamefont {R.}~\bibnamefont {Newman}},
  \bibinfo {author} {\bibfnamefont {R.}~\bibnamefont {Cottis}},\ and\ \bibinfo
  {author} {\bibfnamefont {K.}~\bibnamefont {Sieradzki}},\ }\bibfield  {title}
  {\bibinfo {title} {Validation of a {P}ercolation {M}odel for {P}assivation of
  {F}e-{C}r alloys: {T}wo-{D}imensional {C}omputer {S}imulations},\ }\href@noop
  {} {\bibfield  {journal} {\bibinfo  {journal} {Journal of the Electrochemical
  Society}\ }\textbf {\bibinfo {volume} {137}},\ \bibinfo {pages} {435}
  (\bibinfo {year} {1990})}\BibitemShut {NoStop}%
\bibitem [{\citenamefont {Xie}\ \emph {et~al.}(2021{\natexlab{a}})\citenamefont
  {Xie}, \citenamefont {Artymowicz}, \citenamefont {Lopes}, \citenamefont
  {Aiello}, \citenamefont {Wang}, \citenamefont {Hart}, \citenamefont {Anber},
  \citenamefont {Taheri}, \citenamefont {Zhuang}, \citenamefont {Newman} \emph
  {et~al.}}]{xie2021percolation}%
  \BibitemOpen
  \bibfield  {author} {\bibinfo {author} {\bibfnamefont {Y.}~\bibnamefont
  {Xie}}, \bibinfo {author} {\bibfnamefont {D.~M.}\ \bibnamefont {Artymowicz}},
  \bibinfo {author} {\bibfnamefont {P.~P.}\ \bibnamefont {Lopes}}, \bibinfo
  {author} {\bibfnamefont {A.}~\bibnamefont {Aiello}}, \bibinfo {author}
  {\bibfnamefont {D.}~\bibnamefont {Wang}}, \bibinfo {author} {\bibfnamefont
  {J.~L.}\ \bibnamefont {Hart}}, \bibinfo {author} {\bibfnamefont
  {E.}~\bibnamefont {Anber}}, \bibinfo {author} {\bibfnamefont {M.~L.}\
  \bibnamefont {Taheri}}, \bibinfo {author} {\bibfnamefont {H.}~\bibnamefont
  {Zhuang}}, \bibinfo {author} {\bibfnamefont {R.~C.}\ \bibnamefont {Newman}},
  \emph {et~al.},\ }\bibfield  {title} {\bibinfo {title} {A percolation theory
  for designing corrosion-resistant alloys},\ }\href@noop {} {\bibfield
  {journal} {\bibinfo  {journal} {Nature materials}\ }\textbf {\bibinfo
  {volume} {20}},\ \bibinfo {pages} {789} (\bibinfo {year}
  {2021}{\natexlab{a}})}\BibitemShut {NoStop}%
\bibitem [{\citenamefont {Newman}\ \emph {et~al.}(1988)\citenamefont {Newman},
  \citenamefont {Meng},\ and\ \citenamefont
  {Sieradzki}}]{newman1988validation}%
  \BibitemOpen
  \bibfield  {author} {\bibinfo {author} {\bibfnamefont {R.}~\bibnamefont
  {Newman}}, \bibinfo {author} {\bibfnamefont {F.~T.}\ \bibnamefont {Meng}},\
  and\ \bibinfo {author} {\bibfnamefont {K.}~\bibnamefont {Sieradzki}},\
  }\bibfield  {title} {\bibinfo {title} {Validation of a percolation model for
  passivation of fe-cr alloys: I current efficiency in the incompletely
  passivated state},\ }\href@noop {} {\bibfield  {journal} {\bibinfo  {journal}
  {Corrosion science}\ }\textbf {\bibinfo {volume} {28}},\ \bibinfo {pages}
  {523} (\bibinfo {year} {1988})}\BibitemShut {NoStop}%
\bibitem [{\citenamefont {Wagner}\ \emph {et~al.}(1997)\citenamefont {Wagner},
  \citenamefont {Brankovic}, \citenamefont {Dimitrov},\ and\ \citenamefont
  {Sieradzki}}]{wagner1997dealloying}%
  \BibitemOpen
  \bibfield  {author} {\bibinfo {author} {\bibfnamefont {K.}~\bibnamefont
  {Wagner}}, \bibinfo {author} {\bibfnamefont {S.}~\bibnamefont {Brankovic}},
  \bibinfo {author} {\bibfnamefont {N.}~\bibnamefont {Dimitrov}},\ and\
  \bibinfo {author} {\bibfnamefont {K.}~\bibnamefont {Sieradzki}},\ }\bibfield
  {title} {\bibinfo {title} {Dealloying below the critical potential},\
  }\href@noop {} {\bibfield  {journal} {\bibinfo  {journal} {Journal of the
  Electrochemical Society}\ }\textbf {\bibinfo {volume} {144}},\ \bibinfo
  {pages} {3545} (\bibinfo {year} {1997})}\BibitemShut {NoStop}%
\bibitem [{\citenamefont {Chen}\ \emph {et~al.}(2023)\citenamefont {Chen},
  \citenamefont {Li}, \citenamefont {Zhu},\ and\ \citenamefont
  {Zhuang}}]{chen2023chemical}%
  \BibitemOpen
  \bibfield  {author} {\bibinfo {author} {\bibfnamefont {W.}~\bibnamefont
  {Chen}}, \bibinfo {author} {\bibfnamefont {L.}~\bibnamefont {Li}}, \bibinfo
  {author} {\bibfnamefont {Q.}~\bibnamefont {Zhu}},\ and\ \bibinfo {author}
  {\bibfnamefont {H.}~\bibnamefont {Zhuang}},\ }\bibfield  {title} {\bibinfo
  {title} {Chemical short-range order in complex concentrated alloys},\
  }\href@noop {} {\bibfield  {journal} {\bibinfo  {journal} {MRS Bulletin}\
  }\textbf {\bibinfo {volume} {48}},\ \bibinfo {pages} {762} (\bibinfo {year}
  {2023})}\BibitemShut {NoStop}%
\bibitem [{\citenamefont {Taheri}\ \emph {et~al.}(2023)\citenamefont {Taheri},
  \citenamefont {Anber}, \citenamefont {Barnett}, \citenamefont {Billinge},
  \citenamefont {Birbilis}, \citenamefont {DeCost}, \citenamefont {Foley},
  \citenamefont {Holcombe}, \citenamefont {Hollenbach}, \citenamefont {Joress}
  \emph {et~al.}}]{taheri2023understanding}%
  \BibitemOpen
  \bibfield  {author} {\bibinfo {author} {\bibfnamefont {M.~L.}\ \bibnamefont
  {Taheri}}, \bibinfo {author} {\bibfnamefont {E.}~\bibnamefont {Anber}},
  \bibinfo {author} {\bibfnamefont {A.}~\bibnamefont {Barnett}}, \bibinfo
  {author} {\bibfnamefont {S.}~\bibnamefont {Billinge}}, \bibinfo {author}
  {\bibfnamefont {N.}~\bibnamefont {Birbilis}}, \bibinfo {author}
  {\bibfnamefont {B.}~\bibnamefont {DeCost}}, \bibinfo {author} {\bibfnamefont
  {D.~L.}\ \bibnamefont {Foley}}, \bibinfo {author} {\bibfnamefont
  {E.}~\bibnamefont {Holcombe}}, \bibinfo {author} {\bibfnamefont
  {J.}~\bibnamefont {Hollenbach}}, \bibinfo {author} {\bibfnamefont
  {H.}~\bibnamefont {Joress}}, \emph {et~al.},\ }\bibfield  {title} {\bibinfo
  {title} {Understanding and leveraging short-range order in compositionally
  complex alloys},\ }\href@noop {} {\bibfield  {journal} {\bibinfo  {journal}
  {MRS Bulletin}\ }\textbf {\bibinfo {volume} {48}},\ \bibinfo {pages} {1280}
  (\bibinfo {year} {2023})}\BibitemShut {NoStop}%
\bibitem [{\citenamefont {Yu}(1994)}]{yu1994correlated}%
  \BibitemOpen
  \bibfield  {author} {\bibinfo {author} {\bibfnamefont {G.}~\bibnamefont
  {Yu}},\ }\bibfield  {title} {\bibinfo {title} {Correlated percolation in
  solid solutions with short-range order},\ }\href@noop {} {\bibfield
  {journal} {\bibinfo  {journal} {Philosophical Magazine B}\ }\textbf {\bibinfo
  {volume} {69}},\ \bibinfo {pages} {95} (\bibinfo {year} {1994})}\BibitemShut
  {NoStop}%
\bibitem [{\citenamefont {Cowley}(1960)}]{cowley1960short}%
  \BibitemOpen
  \bibfield  {author} {\bibinfo {author} {\bibfnamefont {J.}~\bibnamefont
  {Cowley}},\ }\bibfield  {title} {\bibinfo {title} {Short-and long-range order
  parameters in disordered solid solutions},\ }\href@noop {} {\bibfield
  {journal} {\bibinfo  {journal} {Physical Review}\ }\textbf {\bibinfo {volume}
  {120}},\ \bibinfo {pages} {1648} (\bibinfo {year} {1960})}\BibitemShut
  {NoStop}%
\bibitem [{\citenamefont {Frary}\ and\ \citenamefont
  {Schuh}(2007)}]{frary2007correlation}%
  \BibitemOpen
  \bibfield  {author} {\bibinfo {author} {\bibfnamefont {M.~E.}\ \bibnamefont
  {Frary}}\ and\ \bibinfo {author} {\bibfnamefont {C.~A.}\ \bibnamefont
  {Schuh}},\ }\bibfield  {title} {\bibinfo {title} {Correlation-space
  description of the percolation transition in composite microstructures},\
  }\href@noop {} {\bibfield  {journal} {\bibinfo  {journal} {Physical Review
  E}\ }\textbf {\bibinfo {volume} {76}},\ \bibinfo {pages} {041108} (\bibinfo
  {year} {2007})}\BibitemShut {NoStop}%
\bibitem [{\citenamefont {Reynolds}\ \emph {et~al.}(1980)\citenamefont
  {Reynolds}, \citenamefont {Stanley},\ and\ \citenamefont
  {Klein}}]{reynolds1980large}%
  \BibitemOpen
  \bibfield  {author} {\bibinfo {author} {\bibfnamefont {P.~J.}\ \bibnamefont
  {Reynolds}}, \bibinfo {author} {\bibfnamefont {H.~E.}\ \bibnamefont
  {Stanley}},\ and\ \bibinfo {author} {\bibfnamefont {W.}~\bibnamefont
  {Klein}},\ }\bibfield  {title} {\bibinfo {title} {Large-cell {M}onte {C}arlo
  renormalization group for percolation},\ }\href@noop {} {\bibfield  {journal}
  {\bibinfo  {journal} {Physical Review B}\ }\textbf {\bibinfo {volume} {21}},\
  \bibinfo {pages} {1223} (\bibinfo {year} {1980})}\BibitemShut {NoStop}%
\bibitem [{\citenamefont {Sotta}\ and\ \citenamefont
  {Long}(2003)}]{sotta2003crossover}%
  \BibitemOpen
  \bibfield  {author} {\bibinfo {author} {\bibfnamefont {P.}~\bibnamefont
  {Sotta}}\ and\ \bibinfo {author} {\bibfnamefont {D.}~\bibnamefont {Long}},\
  }\bibfield  {title} {\bibinfo {title} {The crossover from {2D} to {3D}
  percolation: theory and numerical simulations},\ }\href@noop {} {\bibfield
  {journal} {\bibinfo  {journal} {The European Physical Journal E}\ }\textbf
  {\bibinfo {volume} {11}},\ \bibinfo {pages} {375} (\bibinfo {year}
  {2003})}\BibitemShut {NoStop}%
\bibitem [{\citenamefont {Zekri}\ \emph {et~al.}(2011)\citenamefont {Zekri},
  \citenamefont {Kaiss}, \citenamefont {Clerc}, \citenamefont {Porterie},\ and\
  \citenamefont {Zekri}}]{zekri20112d}%
  \BibitemOpen
  \bibfield  {author} {\bibinfo {author} {\bibfnamefont {L.}~\bibnamefont
  {Zekri}}, \bibinfo {author} {\bibfnamefont {A.}~\bibnamefont {Kaiss}},
  \bibinfo {author} {\bibfnamefont {J.-P.}\ \bibnamefont {Clerc}}, \bibinfo
  {author} {\bibfnamefont {B.}~\bibnamefont {Porterie}},\ and\ \bibinfo
  {author} {\bibfnamefont {N.}~\bibnamefont {Zekri}},\ }\bibfield  {title}
  {\bibinfo {title} {{2D}--to--{3D} percolation crossover of metal--insulator
  composites},\ }\href@noop {} {\bibfield  {journal} {\bibinfo  {journal}
  {Physics Letters A}\ }\textbf {\bibinfo {volume} {375}},\ \bibinfo {pages}
  {346} (\bibinfo {year} {2011})}\BibitemShut {NoStop}%
\bibitem [{\citenamefont {Fey}\ and\ \citenamefont
  {Beyerlein}(2022)}]{fey2022random}%
  \BibitemOpen
  \bibfield  {author} {\bibinfo {author} {\bibfnamefont {L.~T.}\ \bibnamefont
  {Fey}}\ and\ \bibinfo {author} {\bibfnamefont {I.~J.}\ \bibnamefont
  {Beyerlein}},\ }\bibfield  {title} {\bibinfo {title} {Random generation of
  lattice structures with short-range order},\ }\href@noop {} {\bibfield
  {journal} {\bibinfo  {journal} {Integrating Materials and Manufacturing
  Innovation}\ }\textbf {\bibinfo {volume} {11}},\ \bibinfo {pages} {382}
  (\bibinfo {year} {2022})}\BibitemShut {NoStop}%
\bibitem [{\citenamefont {Gehlen}\ and\ \citenamefont
  {Cohen}(1965)}]{gehlen1965computer}%
  \BibitemOpen
  \bibfield  {author} {\bibinfo {author} {\bibfnamefont {P.~t.}\ \bibnamefont
  {Gehlen}}\ and\ \bibinfo {author} {\bibfnamefont {J.}~\bibnamefont {Cohen}},\
  }\bibfield  {title} {\bibinfo {title} {Computer simulation of the structure
  associated with local order in alloys},\ }\href@noop {} {\bibfield  {journal}
  {\bibinfo  {journal} {Physical Review}\ }\textbf {\bibinfo {volume} {139}},\
  \bibinfo {pages} {A844} (\bibinfo {year} {1965})}\BibitemShut {NoStop}%
\bibitem [{\citenamefont {Wolverton}\ \emph {et~al.}(2000)\citenamefont
  {Wolverton}, \citenamefont {Ozolins},\ and\ \citenamefont
  {Zunger}}]{wolverton2000short}%
  \BibitemOpen
  \bibfield  {author} {\bibinfo {author} {\bibfnamefont {C.}~\bibnamefont
  {Wolverton}}, \bibinfo {author} {\bibfnamefont {V.}~\bibnamefont {Ozolins}},\
  and\ \bibinfo {author} {\bibfnamefont {A.}~\bibnamefont {Zunger}},\
  }\bibfield  {title} {\bibinfo {title} {Short-range-order types in binary
  alloys: a reflection of coherent phase stability},\ }\href@noop {} {\bibfield
   {journal} {\bibinfo  {journal} {Journal of Physics: Condensed Matter}\
  }\textbf {\bibinfo {volume} {12}},\ \bibinfo {pages} {2749} (\bibinfo {year}
  {2000})}\BibitemShut {NoStop}%
\bibitem [{\citenamefont {Fantin}\ \emph {et~al.}(2020)\citenamefont {Fantin},
  \citenamefont {Lepore}, \citenamefont {Manzoni}, \citenamefont {Kasatikov},
  \citenamefont {Scherb}, \citenamefont {Huthwelker}, \citenamefont
  {d'Acapito},\ and\ \citenamefont {Schumacher}}]{fantin2020short}%
  \BibitemOpen
  \bibfield  {author} {\bibinfo {author} {\bibfnamefont {A.}~\bibnamefont
  {Fantin}}, \bibinfo {author} {\bibfnamefont {G.~O.}\ \bibnamefont {Lepore}},
  \bibinfo {author} {\bibfnamefont {A.~M.}\ \bibnamefont {Manzoni}}, \bibinfo
  {author} {\bibfnamefont {S.}~\bibnamefont {Kasatikov}}, \bibinfo {author}
  {\bibfnamefont {T.}~\bibnamefont {Scherb}}, \bibinfo {author} {\bibfnamefont
  {T.}~\bibnamefont {Huthwelker}}, \bibinfo {author} {\bibfnamefont
  {F.}~\bibnamefont {d'Acapito}},\ and\ \bibinfo {author} {\bibfnamefont
  {G.}~\bibnamefont {Schumacher}},\ }\bibfield  {title} {\bibinfo {title}
  {Short-range chemical order and local lattice distortion in a compositionally
  complex alloy},\ }\href@noop {} {\bibfield  {journal} {\bibinfo  {journal}
  {Acta Materialia}\ }\textbf {\bibinfo {volume} {193}},\ \bibinfo {pages}
  {329} (\bibinfo {year} {2020})}\BibitemShut {NoStop}%
\bibitem [{\citenamefont {Yildirim}\ and\ \citenamefont {Brown}(2021)}]{rdfpy}%
  \BibitemOpen
  \bibfield  {author} {\bibinfo {author} {\bibfnamefont {B.}~\bibnamefont
  {Yildirim}}\ and\ \bibinfo {author} {\bibfnamefont {H.~G.}\ \bibnamefont
  {Brown}},\ }\href {https://doi.org/10.5281/zenodo.4625675} {\bibinfo {title}
  {by256/rdfpy: rdfpy-v1.0.0}} (\bibinfo {year} {2021})\BibitemShut {NoStop}%
\bibitem [{Sup()}]{Supp}%
  \BibitemOpen
  \href@noop {} {}\bibinfo {note} {See Supplemental Material at [URL will be
  inserted by publisher] for additional computational details, percolation
  rules, and percolation threshold distributions.}\BibitemShut {Stop}%
\bibitem [{\citenamefont {Hoshen}\ and\ \citenamefont
  {Kopelman}(1976)}]{hoshen1976percolation}%
  \BibitemOpen
  \bibfield  {author} {\bibinfo {author} {\bibfnamefont {J.}~\bibnamefont
  {Hoshen}}\ and\ \bibinfo {author} {\bibfnamefont {R.}~\bibnamefont
  {Kopelman}},\ }\bibfield  {title} {\bibinfo {title} {Percolation and cluster
  distribution. {I}. {C}luster multiple labeling technique and critical
  concentration algorithm},\ }\href@noop {} {\bibfield  {journal} {\bibinfo
  {journal} {Physical Review B}\ }\textbf {\bibinfo {volume} {14}},\ \bibinfo
  {pages} {3438} (\bibinfo {year} {1976})}\BibitemShut {NoStop}%
\bibitem [{\citenamefont {Al-Futaisi}\ and\ \citenamefont
  {Patzek}(2003)}]{al2003extension}%
  \BibitemOpen
  \bibfield  {author} {\bibinfo {author} {\bibfnamefont {A.}~\bibnamefont
  {Al-Futaisi}}\ and\ \bibinfo {author} {\bibfnamefont {T.~W.}\ \bibnamefont
  {Patzek}},\ }\bibfield  {title} {\bibinfo {title} {Extension of
  {H}oshen--{K}opelman algorithm to non-lattice environments},\ }\href@noop {}
  {\bibfield  {journal} {\bibinfo  {journal} {Physica A: Statistical Mechanics
  and its Applications}\ }\textbf {\bibinfo {volume} {321}},\ \bibinfo {pages}
  {665} (\bibinfo {year} {2003})}\BibitemShut {NoStop}%
\bibitem [{\citenamefont {Arfken}\ \emph {et~al.}(2011)\citenamefont {Arfken},
  \citenamefont {Weber},\ and\ \citenamefont
  {Harris}}]{arfken2011mathematical}%
  \BibitemOpen
  \bibfield  {author} {\bibinfo {author} {\bibfnamefont {G.~B.}\ \bibnamefont
  {Arfken}}, \bibinfo {author} {\bibfnamefont {H.~J.}\ \bibnamefont {Weber}},\
  and\ \bibinfo {author} {\bibfnamefont {F.~E.}\ \bibnamefont {Harris}},\
  }\href@noop {} {\emph {\bibinfo {title} {Mathematical methods for physicists:
  a comprehensive guide}}}\ (\bibinfo  {publisher} {Academic press},\ \bibinfo
  {year} {2011})\BibitemShut {NoStop}%
\bibitem [{\citenamefont {Lorenz}\ \emph {et~al.}(2000)\citenamefont {Lorenz},
  \citenamefont {May},\ and\ \citenamefont {Ziff}}]{lorenz2000similarity}%
  \BibitemOpen
  \bibfield  {author} {\bibinfo {author} {\bibfnamefont {C.~D.}\ \bibnamefont
  {Lorenz}}, \bibinfo {author} {\bibfnamefont {R.}~\bibnamefont {May}},\ and\
  \bibinfo {author} {\bibfnamefont {R.~M.}\ \bibnamefont {Ziff}},\ }\bibfield
  {title} {\bibinfo {title} {Similarity of percolation thresholds on the {HCP}
  and {FCC} lattices},\ }\href@noop {} {\bibfield  {journal} {\bibinfo
  {journal} {Journal of Statistical Physics}\ }\textbf {\bibinfo {volume}
  {98}},\ \bibinfo {pages} {961} (\bibinfo {year} {2000})}\BibitemShut
  {NoStop}%
\bibitem [{\citenamefont {Cheng}\ \emph {et~al.}(1967)\citenamefont {Cheng},
  \citenamefont {Wynblatt},\ and\ \citenamefont {Dorn}}]{cheng1967vacancy}%
  \BibitemOpen
  \bibfield  {author} {\bibinfo {author} {\bibfnamefont {C.}~\bibnamefont
  {Cheng}}, \bibinfo {author} {\bibfnamefont {P.~P.}\ \bibnamefont
  {Wynblatt}},\ and\ \bibinfo {author} {\bibfnamefont {J.}~\bibnamefont
  {Dorn}},\ }\bibfield  {title} {\bibinfo {title} {Vacancy models for
  concentrated binary alloys—{I} short-range ordered and clustered alloys},\
  }\href@noop {} {\bibfield  {journal} {\bibinfo  {journal} {Acta
  Metallurgica}\ }\textbf {\bibinfo {volume} {15}},\ \bibinfo {pages} {1035}
  (\bibinfo {year} {1967})}\BibitemShut {NoStop}%
\bibitem [{\citenamefont {Zierenberg}\ \emph {et~al.}(2017)\citenamefont
  {Zierenberg}, \citenamefont {Fricke}, \citenamefont {Marenz}, \citenamefont
  {Spitzner}, \citenamefont {Blavatska},\ and\ \citenamefont
  {Janke}}]{zierenberg2017percolation}%
  \BibitemOpen
  \bibfield  {author} {\bibinfo {author} {\bibfnamefont {J.}~\bibnamefont
  {Zierenberg}}, \bibinfo {author} {\bibfnamefont {N.}~\bibnamefont {Fricke}},
  \bibinfo {author} {\bibfnamefont {M.}~\bibnamefont {Marenz}}, \bibinfo
  {author} {\bibfnamefont {F.}~\bibnamefont {Spitzner}}, \bibinfo {author}
  {\bibfnamefont {V.}~\bibnamefont {Blavatska}},\ and\ \bibinfo {author}
  {\bibfnamefont {W.}~\bibnamefont {Janke}},\ }\bibfield  {title} {\bibinfo
  {title} {Percolation thresholds and fractal dimensions for square and cubic
  lattices with long-range correlated defects},\ }\href@noop {} {\bibfield
  {journal} {\bibinfo  {journal} {Physical Review E}\ }\textbf {\bibinfo
  {volume} {96}},\ \bibinfo {pages} {062125} (\bibinfo {year}
  {2017})}\BibitemShut {NoStop}%
\bibitem [{\citenamefont {Weinrib}(1984)}]{weinrib1984long}%
  \BibitemOpen
  \bibfield  {author} {\bibinfo {author} {\bibfnamefont {A.}~\bibnamefont
  {Weinrib}},\ }\bibfield  {title} {\bibinfo {title} {Long-range correlated
  percolation},\ }\href@noop {} {\bibfield  {journal} {\bibinfo  {journal}
  {Physical Review B}\ }\textbf {\bibinfo {volume} {29}},\ \bibinfo {pages}
  {387} (\bibinfo {year} {1984})}\BibitemShut {NoStop}%
\bibitem [{\citenamefont {Lorenz}\ and\ \citenamefont
  {Ziff}(1998)}]{lorenz1998precise}%
  \BibitemOpen
  \bibfield  {author} {\bibinfo {author} {\bibfnamefont {C.~D.}\ \bibnamefont
  {Lorenz}}\ and\ \bibinfo {author} {\bibfnamefont {R.~M.}\ \bibnamefont
  {Ziff}},\ }\bibfield  {title} {\bibinfo {title} {Precise determination of the
  bond percolation thresholds and finite-size scaling corrections for the sc,
  fcc, and bcc lattices},\ }\href@noop {} {\bibfield  {journal} {\bibinfo
  {journal} {Physical Review E}\ }\textbf {\bibinfo {volume} {57}},\ \bibinfo
  {pages} {230} (\bibinfo {year} {1998})}\BibitemShut {NoStop}%
\bibitem [{\citenamefont {Ding}\ \emph {et~al.}(2014)\citenamefont {Ding},
  \citenamefont {Li}, \citenamefont {Zhang}, \citenamefont {Lu},\ and\
  \citenamefont {Ji}}]{ding2014numerical}%
  \BibitemOpen
  \bibfield  {author} {\bibinfo {author} {\bibfnamefont {B.}~\bibnamefont
  {Ding}}, \bibinfo {author} {\bibfnamefont {C.}~\bibnamefont {Li}}, \bibinfo
  {author} {\bibfnamefont {M.}~\bibnamefont {Zhang}}, \bibinfo {author}
  {\bibfnamefont {G.}~\bibnamefont {Lu}},\ and\ \bibinfo {author}
  {\bibfnamefont {F.}~\bibnamefont {Ji}},\ }\bibfield  {title} {\bibinfo
  {title} {Numerical analysis of percolation cluster size distribution in
  two-dimensional and three-dimensional lattices},\ }\href@noop {} {\bibfield
  {journal} {\bibinfo  {journal} {The European Physical Journal B}\ }\textbf
  {\bibinfo {volume} {87}},\ \bibinfo {pages} {1} (\bibinfo {year}
  {2014})}\BibitemShut {NoStop}%
\bibitem [{\citenamefont {N{\'o}brega}\ and\ \citenamefont
  {Stauffer}(2003)}]{nobrega2003corrections}%
  \BibitemOpen
  \bibfield  {author} {\bibinfo {author} {\bibfnamefont {R.}~\bibnamefont
  {N{\'o}brega}}\ and\ \bibinfo {author} {\bibfnamefont {D.}~\bibnamefont
  {Stauffer}},\ }\bibfield  {title} {\bibinfo {title} {Corrections to finite
  size scaling in percolation},\ }\href@noop {} {\bibfield  {journal} {\bibinfo
   {journal} {Brazilian Journal of Physics}\ }\textbf {\bibinfo {volume}
  {33}},\ \bibinfo {pages} {616} (\bibinfo {year} {2003})}\BibitemShut
  {NoStop}%
\bibitem [{\citenamefont {Ziff}\ and\ \citenamefont
  {Newman}(2002)}]{ziff2002convergence}%
  \BibitemOpen
  \bibfield  {author} {\bibinfo {author} {\bibfnamefont {R.~M.}\ \bibnamefont
  {Ziff}}\ and\ \bibinfo {author} {\bibfnamefont {M.}~\bibnamefont {Newman}},\
  }\bibfield  {title} {\bibinfo {title} {Convergence of threshold estimates for
  two-dimensional percolation},\ }\href@noop {} {\bibfield  {journal} {\bibinfo
   {journal} {Physical Review E}\ }\textbf {\bibinfo {volume} {66}},\ \bibinfo
  {pages} {016129} (\bibinfo {year} {2002})}\BibitemShut {NoStop}%
\bibitem [{\citenamefont {Xie}\ \emph {et~al.}(2021{\natexlab{b}})\citenamefont
  {Xie}, \citenamefont {Chatterjee}, \citenamefont {Liu}, \citenamefont {Jin},\
  and\ \citenamefont {Sieradzki}}]{xie2021passivation}%
  \BibitemOpen
  \bibfield  {author} {\bibinfo {author} {\bibfnamefont {Y.}~\bibnamefont
  {Xie}}, \bibinfo {author} {\bibfnamefont {S.}~\bibnamefont {Chatterjee}},
  \bibinfo {author} {\bibfnamefont {L.-Z.}\ \bibnamefont {Liu}}, \bibinfo
  {author} {\bibfnamefont {H.-J.}\ \bibnamefont {Jin}},\ and\ \bibinfo {author}
  {\bibfnamefont {K.}~\bibnamefont {Sieradzki}},\ }\bibfield  {title} {\bibinfo
  {title} {Passivation of {C}u-{R}h {A}lloys},\ }\href@noop {} {\bibfield
  {journal} {\bibinfo  {journal} {Journal of The Electrochemical Society}\
  }\textbf {\bibinfo {volume} {168}},\ \bibinfo {pages} {071505} (\bibinfo
  {year} {2021}{\natexlab{b}})}\BibitemShut {NoStop}%
\bibitem [{\citenamefont {Walsh}\ \emph {et~al.}(2023)\citenamefont {Walsh},
  \citenamefont {Abu-Odeh},\ and\ \citenamefont
  {Asta}}]{walsh2023reconsidering}%
  \BibitemOpen
  \bibfield  {author} {\bibinfo {author} {\bibfnamefont {F.}~\bibnamefont
  {Walsh}}, \bibinfo {author} {\bibfnamefont {A.}~\bibnamefont {Abu-Odeh}},\
  and\ \bibinfo {author} {\bibfnamefont {M.}~\bibnamefont {Asta}},\ }\bibfield
  {title} {\bibinfo {title} {Reconsidering short-range order in complex
  concentrated alloys},\ }\href@noop {} {\bibfield  {journal} {\bibinfo
  {journal} {MRS Bulletin}\ }\textbf {\bibinfo {volume} {48}},\ \bibinfo
  {pages} {753} (\bibinfo {year} {2023})}\BibitemShut {NoStop}%
\bibitem [{\citenamefont {Kadkhodaei}\ and\ \citenamefont
  {Mu{\~n}oz}(2021)}]{kadkhodaei2021cluster}%
  \BibitemOpen
  \bibfield  {author} {\bibinfo {author} {\bibfnamefont {S.}~\bibnamefont
  {Kadkhodaei}}\ and\ \bibinfo {author} {\bibfnamefont {J.~A.}\ \bibnamefont
  {Mu{\~n}oz}},\ }\bibfield  {title} {\bibinfo {title} {Cluster expansion of
  alloy theory: a review of historical development and modern innovations},\
  }\href@noop {} {\bibfield  {journal} {\bibinfo  {journal} {JOM}\ }\textbf
  {\bibinfo {volume} {73}},\ \bibinfo {pages} {3326} (\bibinfo {year}
  {2021})}\BibitemShut {NoStop}%
\bibitem [{\citenamefont {Berchenko}\ \emph {et~al.}(2009)\citenamefont
  {Berchenko}, \citenamefont {Artzy-Randrup}, \citenamefont {Teicher},\ and\
  \citenamefont {Stone}}]{berchenko2009emergence}%
  \BibitemOpen
  \bibfield  {author} {\bibinfo {author} {\bibfnamefont {Y.}~\bibnamefont
  {Berchenko}}, \bibinfo {author} {\bibfnamefont {Y.}~\bibnamefont
  {Artzy-Randrup}}, \bibinfo {author} {\bibfnamefont {M.}~\bibnamefont
  {Teicher}},\ and\ \bibinfo {author} {\bibfnamefont {L.}~\bibnamefont
  {Stone}},\ }\bibfield  {title} {\bibinfo {title} {Emergence and size of the
  giant component in clustered random graphs with a given degree
  distribution},\ }\href@noop {} {\bibfield  {journal} {\bibinfo  {journal}
  {Physical review letters}\ }\textbf {\bibinfo {volume} {102}},\ \bibinfo
  {pages} {138701} (\bibinfo {year} {2009})}\BibitemShut {NoStop}%
\bibitem [{\citenamefont {Tian}\ and\ \citenamefont
  {Sun}(2022)}]{tian2022percolation}%
  \BibitemOpen
  \bibfield  {author} {\bibinfo {author} {\bibfnamefont {Y.}~\bibnamefont
  {Tian}}\ and\ \bibinfo {author} {\bibfnamefont {P.}~\bibnamefont {Sun}},\
  }\bibfield  {title} {\bibinfo {title} {Percolation may explain efficiency,
  robustness, and economy of the brain},\ }\href@noop {} {\bibfield  {journal}
  {\bibinfo  {journal} {Network Neuroscience}\ }\textbf {\bibinfo {volume}
  {6}},\ \bibinfo {pages} {765} (\bibinfo {year} {2022})}\BibitemShut {NoStop}%
\bibitem [{\citenamefont {Yang}\ \emph {et~al.}(2023)\citenamefont {Yang},
  \citenamefont {Zhou}, \citenamefont {Yin}, \citenamefont {Wang},
  \citenamefont {Yu}, \citenamefont {Olszta}, \citenamefont {Zhang},
  \citenamefont {Zeltmann}, \citenamefont {Li}, \citenamefont {Jin} \emph
  {et~al.}}]{yang2023one}%
  \BibitemOpen
  \bibfield  {author} {\bibinfo {author} {\bibfnamefont {Y.}~\bibnamefont
  {Yang}}, \bibinfo {author} {\bibfnamefont {W.}~\bibnamefont {Zhou}}, \bibinfo
  {author} {\bibfnamefont {S.}~\bibnamefont {Yin}}, \bibinfo {author}
  {\bibfnamefont {S.~Y.}\ \bibnamefont {Wang}}, \bibinfo {author}
  {\bibfnamefont {Q.}~\bibnamefont {Yu}}, \bibinfo {author} {\bibfnamefont
  {M.~J.}\ \bibnamefont {Olszta}}, \bibinfo {author} {\bibfnamefont {Y.-Q.}\
  \bibnamefont {Zhang}}, \bibinfo {author} {\bibfnamefont {S.~E.}\ \bibnamefont
  {Zeltmann}}, \bibinfo {author} {\bibfnamefont {M.}~\bibnamefont {Li}},
  \bibinfo {author} {\bibfnamefont {M.}~\bibnamefont {Jin}}, \emph {et~al.},\
  }\bibfield  {title} {\bibinfo {title} {One dimensional wormhole corrosion in
  metals},\ }\href@noop {} {\bibfield  {journal} {\bibinfo  {journal} {Nature
  Communications}\ }\textbf {\bibinfo {volume} {14}},\ \bibinfo {pages} {988}
  (\bibinfo {year} {2023})}\BibitemShut {NoStop}%
\end{thebibliography}%
\end{document}